\documentclass[a4paper,12pt,english]{article}

\usepackage{helvet}         
\usepackage{courier}        
\usepackage{type1cm}        
\usepackage{makeidx}         
\usepackage{graphicx}        
\usepackage{multicol}        
\usepackage[bottom]{footmisc}

\usepackage{cite}
\usepackage{mathtools}
\usepackage{empheq}

\usepackage[mathscr]{euscript}
\usepackage[all,color]{xy}
\usepackage{relsize}
\usepackage{url}
\usepackage{cancel}
\usepackage{caption}
\usepackage[T1]{fontenc}
\usepackage{color}
\usepackage{mathtools}

\newcommand{\bbm}{\left(\begin{matrix}}
    \newcommand{\ebm}{\end{matrix}\right)}
\newcommand{\beq}{\begin{eqnarray}}
\newcommand{\eeq}{\end{eqnarray}}

\makeatother

 \def\one{\mbox{1 \kern-.59em {\rm l}}}

\begin{document}
\begin{flushright}
RBI-ThPhys-2021-16
\end{flushright}

\begingroup
{\let\newpage\relax
\title{Four-Dimensional Gravity on a Covariant Noncommutative Space (II)}

\author{G. Manolakos\textsuperscript{1,2},\,P. Manousselis\textsuperscript{1},\,G. Zoupanos\textsuperscript{1,3,4,5}}\date{}
\maketitle}
\begin{center}
\emph{E-mails: gmanol@central.ntua.gr\,, pman@central.ntua.gr\,, George.Zoupanos@cern.ch }
\end{center}

\begin{center}
\itshape\textsuperscript{1}Physics Department, National Technical
University, Athens, Greece\\
\itshape\textsuperscript{2} Division of Theoretical Physics, Ruđer Bošković Institute, Zagreb, Croatia\\
\itshape\textsuperscript{3} Theory Department, CERN\\
\itshape\textsuperscript{4} Max-Planck Institut f\"ur Physik, M\"unchen, Germany\\
\itshape\textsuperscript{5} Institut f\"ur Theoretische Physik der Universit\"at Heidelberg, Germany
\end{center}
\vspace{0.1cm} \emph{Keywords}: gauge theories, four-dimensional gravity, noncommutative spaces, fuzzy de Sitter, noncommutative gravity, spontaneous symmetry breaking 

\maketitle

\abstract{
Based on the construction of the 4-dim noncommutative gravity model described in our previous work, first, a more extended description of the covariant noncommutative space (fuzzy 4-dim de Sitter space), which accommodates the gravity model, is presented and then the corresponding field equations, which are obtained after variation of the previously proposed action, are extracted. Also, a spontaneous breaking of the initial symmetry is performed, this time induced by the introduction of an auxiliary scalar field, and its implications in the reduced theory, which is produced after considering the commutative limit, are examined.}

\maketitle

\section{Introduction}
Besides the traditional, geometrical description of various gravitational theories, there exists an alternative way of approaching them, that is via a gauge-theoretic formulation. The pioneer work towards this direction was made by Utiyama \cite{Utiyama:1956sy} and then many others picked up the torch developing around the original idea \cite{Witten:1988hc,Kibble:1961ba,Stelle:1979aj,MacDowell:1977jt,Ivanov:1980tw,Kibble:1985sn,Kaku:1977pa, Fradkin:1985am,vanproeyen,cham-thesis,Chamseddine:1976bf}. In particular, some indicative results of this development are i) the equivalence of the 3-dim Einstein gravity to a Chern-Simons gauge theory \cite{Witten:1988hc}, ii) the description of the 4-dim Einstein gravity as a gauge theory which admits a spontaneous symmetry breaking with the involvement of an auxiliary scalar field \cite{Kibble:1961ba,Stelle:1979aj} and iii) the description of the 4-dim Weyl gravity as a gauge theory of the conformal group which is broken after the imposition of constraint equations 
\cite{MacDowell:1977jt,Ivanov:1980tw,Kibble:1985sn,Kaku:1977pa,
Fradkin:1985am,vanproeyen,cham-thesis,Chamseddine:1976bf}\footnote{Also, in refs \cite{Oda:2019iwc, Ghilencea:2019jux}, further breaking patterns of the Weyl conformal gravity are studied.}. 

The above, successful, alternative description of gravity theories as gauge theories can be translated to the noncommutative framework (for details on noncommutative geometry see \cite{Connes:1994yd,Madore:2000aq,Szabo:2001kg,Hinchliffe:2002km}). At a technical level, such a translation is suggested by the existence of a well-defined formulation of noncommutative gauge theories \cite{Madore:2000en,Jurco:2000ja,Jurco:2001rq}. However, from a more physical point of view, the whole translation of the above programme to the noncommutative regime is highly-motivated when it comes to the description of the gravitational interaction in case that the notion of spacetime, as it is classically perceived, collapses to a configuration in which the corresponding coordinates cease to commute and therefore the notion of points becomes meaningless. An appropriate candidate for such a configuration is the form of the spacetime considered at the Planck scale, in which a minimal (Planck) length is involved
and thus the spacetime loses its continuity and becomes discretized. In this special case, noncommutative spaces could serve as background spaces on which the description of gravity at Planck scale could be nicely realized and the formulation of such noncommutative gravity theories could be addressed in a customized and well-established gauge-theoretic way, as suggested by the existence of the corresponding approach of ordinary gravity theories, as mentioned above.

Towards this direction, many remarkable contributions have been made following two different approaches within the framework of noncommutativity, due to the fact that noncommutativity can be manifested either by working with ordinary functions and concentrate the feature of noncommutativity into a variation of the multiplication operation that is called star-product, or by considering that the elements of the theory are operators and therefore potentially represented by matrices so that the noncommutative nature is already inherited in the ordinary matrix product, which is inherently noncommutative. In particular, some representative noncommutative field theories  which are formulated employing the star-product and the Seiberg-Witten map \cite{Seiberg:1999vs} (see also \cite{Brandt:2003fx}) can be found in refs \cite{Aschieri:2002mc,Aschieri:2005yw,Aschieri:2005zs,Chamseddine:2000si,Chamseddine:2002fd,Chamseddine:2003we,Aschieri:2009ky,Aschieri:2009mc,Ciric:2016isg,Cacciatori:2002gq,Cacciatori:2002ib,Aschieri:2014xka,Banados:2001xw}, while some indicative publications in which the matrix treatment of noncommutativity (matrix geometries) is adopted leading to matrix models can be found in refs \cite{Banks:1996vh,Ishibashi:1996xs,Aoki:1998vn,Hanada:2005vr,Furuta:2006kk}. For an alternative approach see also refs \cite{Buric:2006di,Buric:2007zx,Buric:2007hb}, in which it is described that the degrees of freedom of the resulting gravitational theory are associated to those of the noncommutative structure. Last, a very recent and systematic approach on constructing noncommutative field theories using braided $L_\infty$-algebras can be found in refs \cite{Ciric:2020eab,Ciric:2021rhi}.

From our perspective, we are oriented towards the matrix-realized models focusing on the specific class of covariant noncommutative spaces\cite{Snyder:1946qz,Yang:1947ud,Madore:1991bw,Grosse:1993uq,Buric:2015wta,Buric:2017yes,Heckman:2014xha}, which are suitable for constructing noncommutative models \cite{Dolan:2001gn,OConnor:2006iny,Medina:2002pc,Medina:2012cs,Gere:2013uaa,Vitale:2012dz,Vitale:2014hca}, especially gravitational ones \cite{Kimura:2002nq,Steinacker:2016vgf,Sperling:2017dts,Sperling:2017gmy,Sperling:2018xrm,Jurman:2013ota,Yang:2006dk} since they are equipped with the property of preserving Lorentz covariance. Along these lines, our first work in the subject was realized in three dimensions \cite{Chatzistavrakidis:2018vfi} (see also \cite{Manolakos:2018hvn}). In particular, in order to translate the 3-dim gauge-theoretic description of gravity to the noncommutative regime, we examined both the Euclidean and Lorentzian cases by considering the covariant noncommutative background spaces to be the $\mathbf{R}_\lambda^3$ \cite{Hammou:2001cc,Gere:2013uaa,Vitale:2012dz,Vitale:2014hca}, which is a foliation of the 3-dim Euclidean space by multiple fuzzy spheres \cite{Madore:1991bw} of different radii and $\mathbf{R}_{\lambda}^{1,2}$ which is the foliation of the 3-dim Minkowski spacetime by adjacent fuzzy hyperboloids \cite{Jurman:2013ota}, respectively. Starting with their isometry groups and working with the well-defined noncommutative gauge-theoretic construction, we produced the two gravity models, in which we calculated the transformations of the various gauge fields and the expressions of their corresponding curvature tensors. Eventually, we proposed an action of Chern-Simons type, found the equations of motion and examined the commutative limit, in which the above results reduced to the expected ones for the 3-dim Einstein gravity.    

Next, aiming at the construction of a 4-dim noncommutative gravity model, we first got involved with the construction of a 4-dim covariant noncommutative space that would serve as the background space on which the gauge theory would be constructed \cite{Manolakos:2019fle} (see also \cite{Manolakos:2019fek}). Motivated by refs \cite{Buric:2015wta,Buric:2017yes,Heckman:2014xha}, we focused on the construction of a covariant fuzzy version of the 4-dim de Sitter space and obtained the defining commutation relation of its coordinates. Then, employing this construction, we moved on with determining the gauge group, which was an extension of the isometry group of the background space, as expected, and obtained the various associated resulting expressions of the necessary quantities. Then, we wrote down a topological action, quadratic on the curvature 2-form and performed an explicit symmetry breaking by imposing specific and motivated constraints. 

In the current work we complete the picture of the above, 4-dim construction. First, we give some more information about the construction of the covariant noncommutative space and show that the given form of 4-dim description is derived from a 2-step decomposition of the initial (enlarged) isometry group. We emphasize in the intermediate first step which manifests the relation of this space to its commutative origin being described in the familiar picture, that is as embedded in the 5-dim Minkowski spacetime. Then, revisiting the action defined in our previous work, we argue on a more motivated way of deriving it, we explicitly show that the background space is indeed a solution of the theory and then we extract the dynamical field equations of the theory by considering fluctuations around the vacuum solution. Next, regarding the symmetry breaking that was realized by the imposition of constraints in our previous work, 
it is now performed in such a way that is spontaneously induced by the presence of an auxiliary scalar field. Finally, we project our results to the commutative limit, which is actually the low-energy (large-scale) regime, and examine and comment on the picture of the gravitational theory with which we eventually result.    

\section{A noncommutative version of the 4-dim de Sitter space}

Our aim is to revisit the formulation of the 4-dim covariant spacetime we introduced in our previous work \cite{Manolakos:2019fle}, which is a fuzzy version of the 4-dim de Sitter space\footnote{Our construction was motivated by the different approaches of \cite{Buric:2015wta, Buric:2017yes} and \cite{Heckman:2014xha} on the formulation of this space.}. We present a 2-step procedure, in which in the first step the space is manifested as an embedding in the 5-dim Minkowski spacetime (with similar arguments to those of the formulation of the fuzzy 2-sphere \cite{Madore:1991bw}), while in the second step we recover the pure 4-dim configuration we presented in our previous work, that is the one we employed for the construction of our gravity model. 

Let us first recall some basic information about the ordinary  (commutative) $dS_4$ spacetime. In general, it consists a maximally symmetric Lorentzian (non-compact) manifold with constant positive curvature and can be considered as the 4-dim Lorentzian analogue of the 4-dim sphere. According to this analogy, in the same manner that the 4-sphere is embedded into the 5-dim Euclidean space, the $dS_4$ admits a description as an embedding into the 5-dim Minkowski spacetime with metric $\eta_{MN}=\mathrm{diag}(-1,1,1,1,1)$, namely:
\begin{equation}
    \eta_{MN}x^M x^N=r^2\,,
\end{equation}
where $r$ is the radius of the space and $M, N=0,\ldots,4$. Last, the isometry group of this spacetime is the $SO(1,4)$ (the Lorentz group of the 5-dim Minkowski spacetime). 

Now, we want to translate the above space to the noncommutative regime. In order to achieve this, according to the fuzzy sphere case, we consider the isometry group of the de Sitter space, $SO(1,4)$ and intend to identify the coordinates and the tensor of noncommutativity with generators of the group. However, in this $dS_4$ case, the above scheme cannot be realized because the covariance of the space is not preserved, that is, if we correspond five of the generators to the coordinates of the spacetime, then the rest of the generators do not suffice to form a Lorentz subgroup, under which the coordinates should transform as vectors. Having already pointed out the importance of covariance, in order to preserve it, we need to minimally extend the considered group to the $SO(1,5)$, in which the coordinates, Lorentz transformations and other operators can be nicely included. Therefore, starting with $SO(1,5)$, we perform a \emph{2-step decomposition}, $SO(1,5)\supset SO(1,4)\supset SO(1,3)$ (maximal subgroups) in order, in the first step, to obtain an explicit configuration of the space which will be easily recognized as the fuzzy version of the $dS_4$ (see \cite{Kimura:2002nq} and \cite{Steinacker:2016vgf,Sperling:2017dts,Sperling:2017gmy,Sperling:2018xrm}) and, in the second step, to result with a 4-dim formulation of the space according to \cite{Snyder:1946qz,Yang:1947ud}. \\

\noindent \underline{\emph{First step}: $SO(1,5)\supset SO(1,4)$}\\\\
We start with the $SO(1,5)$ group which comprises of fifteen (antisymmetric) generators $J_{MN}$,  where $M,N=0,\ldots,5$, which are considered to be hermitian. The commutation relation of the generators of the group is:
\begin{equation}
[J_{MN}, J_{P\Sigma} ] = i(\eta_{MP}J_{N\Sigma} + \eta_{N\Sigma}J_{MP} - \eta_{NP}J_{M\Sigma} - \eta_{M\Sigma}J_{NP} )\,,
\end{equation}
where $\eta_{MN}=\mathrm{diag}(-1,1,1,1,1,1)$ is the corresponding 6-dim Minkowski metric. The first part of the decomposition, $SO(1,5)\supset SO(1,4)$, turns the above commutation relation to the following three:
\begin{itemize}
    \item[-] For $M=m, N=m, P=r, \Sigma=s$:\\
    $[J_{mn},J_{rs}]=i(\eta_{mr}J_{ns}+\eta_{ns}J_{mr}-\eta_{nr}J_{ms}-\eta_{ms}J_{nr})\,.$
    \item[-] For $M=m, N=5, P=r, \Sigma=5$:
            $[J_{m5},J_{n5}]=iJ_{mn}\,$.
    \item[-] For $M=m, N=n, P=r, \Sigma=5$:
        $[J_{mn},J_{r5}]=i(\eta_{mr}J_{n5}-\eta_{nr} J_{m5})\,,$\\ where $m,n,r,s=0,\ldots,4$
        \end{itemize}
    In all above three relations, $\eta_{mn}=\mathrm{diag}(-1,1,1,1,1)$, that is the 5-dim Minkowski metric. 
     In order to convert the generators to physical quantities, we set $\Theta_{mn}\equiv \hbar J_{mn}\,, X_{m}\equiv \lambda J_{m5}$, where $\lambda$ is a parameter of dimension of length. According to these definitions, the above commutation relations become:
    \begin{align}
        [\Theta_{mn},\Theta_{rs}]&=i\hbar(\eta_{mr}\Theta_{ns}+\eta_{ns}\Theta_{mr}-\eta_{nr}\Theta_{ms}-\eta_{ms}\Theta_{nr})\,,\label{firstcomrelfirststep}\\
        [\Theta_{mn},X_{r}]&=i\hbar(\eta_{mr}X_n-\eta_{nr}X_m)\,,\label{secondcomrelfirststep}\\
        [X_m,X_n]&=i\frac{\lambda^2}{\hbar}\Theta_{mn}\,.\label{thirdcomrelfirststep}
    \end{align}
     Assuming that the $SO(1,5)$ generators live in an arbitrary $N$-dim irreducible representation, then the corresponding three independent Casimir invariants \cite{Wybourne:1974,Mountain:1998}, (equal to the rank of the group) are considered as follows:
     \begin{itemize}
    \item[-] The quadratic Casimir:
    \begin{align}
        C_2^{SO(1,5)}&=-\frac{1}{2}\mathrm{Tr}J^2=\frac{1}{2}J_{MN}J^{MN}=\frac{1}{2}J_{mn}J^{mn}+J_{m5}J^{m5}\Rightarrow\nonumber\\
        X^m X_m&=\lambda^2(C_2^{SO(1,5)}-C_2^{SO(1,4)})\equiv r^2\,.\label{embeddingrelation}
    \end{align}
     Since the irreducible representation of the $J_{MN}$ generators of $SO(1,5)$ is a (high) $N$-dim, this means that they can be written down as $N\times N$ matrices. Thus, in order that we legitimately write the trace $\frac{1}{2}J_{mn}J^{mn}$, which lives in the same vector space, as the $SO(1,4)$ quadratic Casimir, namely $C_2^{SO(1,4)}$, we have implicitly assumed that the decomposition $SO(1,5)\supset SO(1,4)$ leads to $SO(1,4)$ irreducible representations of the same dimension. Therefore, we conclude that only dimension-preserving decompositions are allowed and, thus, the representation of the $SO(1,4)$ group after the above decomposition will be also $N$-dim. Also, we have identified $r^2\equiv\lambda^2(C_2^{SO(1,5)}-C_2^{SO(1,4)})$, which is a quantity proportional to the $I_N$ matrix, its specific value is determined by the representations of the groups and takes over the role of the constant (and positive) radius of the manifold\footnote{In case we were aiming at the fuzzy 4-sphere through our 2-step decomposition, the first maximal decomposition would be that of the corresponding compact groups, i.e. $SO(6)\supset SO(5)$. In that case, we would have given the explicit expression of the radius in terms of the representations:\[X^m X_m=\lambda^2(C_2^{SO(6)}-C_2^{SO(5)})=\lambda^2(L(L+4)\mathbf{I}_L-\ell(\ell+3)\mathbf{I}_\ell)\stackrel{L=\ell}{=}\lambda^2L\mathbf{I}_L\equiv r^2\mathbf{I}_L\,,\] where the identification $r^2\equiv\lambda^2L$ has been made. The $r^2$ would be interpreted as the radius of the 4-sphere in which $L$ would be the initial representation of the $SO(6)$ generators and $\ell$ would be the representation of the $SO(5)$ ones, assuming again that $L=\ell$. The above expression of the radius was captured by the general relation for an $SO(d)$ group of the quadratic Casimir, that is $C_2^{SO(d)}=L(L+d-2)$, where $L$ is the spin representation of the corresponding generators. Also, in the commutative limit, that is $L\rightarrow\infty$ and $\lambda\rightarrow 0$, the ordinary $S_4$ is recovered. For more details on the construction of fuzzy 4-sphere see \cite{Steinacker:2016vgf,Sperling:2017dts,Sperling:2017gmy}. The whole discussion for the non-compact groups we are using is more subtle and that is why we postpone it for future work.}.
    \item[-] The cubic $P$ Casimir:
    \begin{align}
        P^{SO(1,5)}&=\epsilon_{MNP\Sigma K\Lambda}J^{MN}J^{P\Sigma}J^{K\Lambda}\nonumber\\
        &=\frac{2}{\hbar^2\lambda}\epsilon_{mnrsq}\left(\Theta^{mn}\{\Theta^{rs},X^q\}+\frac{1}{2}X^m\{\Theta^{nr},\Theta^{sq}\}\right)
    \end{align}
    \item[-] The quartic Casimir:
    \begin{align}
        C_4^{SO(1,5)}&=-\frac{1}{4}\mathrm{Tr}J^4=-\frac{1}{4}J^{MN}J_{NP}J^{P\Sigma}J_{\Sigma M}\Rightarrow\nonumber\\
        C_4^{SO(1,5)}+\frac{r^4}{2\lambda^4}&=-\frac{1}{4\hbar^4}\Theta^{mn}\Theta_{nr}\Theta^{rs}\Theta_{sm}+\frac{1}{4\hbar^2\lambda^2}\left(\{\Theta^{mn}\Theta_{nr},X^r X_m\}+\{\Theta^{mn}X_n,X^r\Theta_{rm}\}\right)\,.
    \end{align}
    Thus, the above cubic and quartic Casimir operators produce relations involving commutators and anticommutators of the operators of the algebra. These relations do not provide any more constraint equations on the space coordinates, rather relate the various operators of the theory in a non-trivial way. 
    
    \item[-] Therefore, we result with a covariant fuzzy version of the 4-dim de Sitter space, which is formulated as an embedding in the 5-dim Minkowski space with embedding equation $X^m X_m=r^2$, where the radius $r$ depends on the representation of the operators and is a $N\times N$ matrix. The feature of noncommutativity is expressed via the relation $[X_m,X_n]=\dfrac{i\lambda^2}{\hbar}\Theta_{mn}$, where $m,n=0,\ldots,4$ and the isometry group is the $SO(1,4)$, as expected since it consists a fuzzy version of the $dS_4$ space. 
\end{itemize}
\noindent \underline{\emph{Second step}: $SO(1,4)\supset SO(1,3)$}\\\\
In this second step we want to express the above 4-dim fuzzy de Sitter space in an $SO(1,3)$ language. This way we expect that we will end up with the space as defined in our previous paper \cite{Manolakos:2019fle}. In order to achieve this, we work as follows:
\begin{itemize}
    \item From eq.\eqref{firstcomrelfirststep}, we have:
    \begin{itemize}
        \item For $m=\mu, n=\nu, r=\rho, s=\sigma$, where $\mu,\nu,\rho,\sigma=0,\ldots, 3$:
        \begin{equation}
            [\Theta_{\mu\nu},\Theta_{\rho\sigma}]=i\hbar\left(\eta_{\mu\rho}\Theta_{\nu\sigma}+\eta_{\nu\sigma}\Theta_{\mu\rho}-\eta_{\nu\rho}\Theta_{\mu\sigma}-\eta_{\mu\sigma}\Theta_{\nu\rho}\right)\,.
        \end{equation}
        \item For $m=\mu, n=\nu, r=\rho, s=4$:
        \begin{equation}
            [\Theta_{\mu\nu},\Theta_{\rho 4}]=i\hbar(\eta_{\mu\rho}\Theta_{\nu 4}-\eta_{\nu\rho}\Theta_{\mu 4})\,.    
        \end{equation}
        \item For $m=\mu, n=4, r=\rho, s=4$:
        \begin{equation}
            [\Theta_{\mu 4},\Theta_{\rho 4}]=i\hbar\Theta_{\mu\rho}\,.
        \end{equation}
    \end{itemize}
    \item From eq.\eqref{secondcomrelfirststep}, we have:
    \begin{itemize}
        \item For $m=\mu, n=\nu, r=\rho$:
        \begin{equation}
            [\Theta_{\mu\nu},X_\rho]=i\hbar(\eta_{\mu\rho}X_\nu-\eta_{\nu\rho}X_\mu)\,.
        \end{equation}
        \item For $m=\mu, n=\nu, r=4$:
        \begin{equation}
            [\Theta_{\mu\nu},X_4]=0\,.
        \end{equation}
        \item For $m=\mu, n=4, r=4$:
        \begin{equation}
            [\Theta_{\mu 4},X_4]=-i\hbar X_\mu\,.
        \end{equation}
        \item For $m=\mu, n=4, r=\rho$:
        \begin{equation}
            [\Theta_{\mu 4},X_\rho]=i\hbar\eta_{\mu\rho}X_4\,.
        \end{equation}
    \end{itemize}
    \item From eq.\eqref{thirdcomrelfirststep}, we have:
    \begin{itemize}
        \item For $m=\mu, n=\nu$:
        \begin{equation}
            [X_\mu,X_\nu]=i\frac{\lambda^2}{\hbar}\Theta_{\mu\nu}\,.
        \end{equation}
        \item For $m=\mu, n=4$:
        \begin{equation}
            [X_\mu,X_4]=i\frac{\lambda^2}{\hbar}\Theta_{\mu 4}\,.
        \end{equation}
    \end{itemize}
    \item We set: $P_\mu\equiv\frac{1}{\lambda}\Theta_{\mu 4}, h\equiv\frac{1}{\lambda}X_4$. Thus, the commutation relations regarding all the operators $\Theta_{\mu\nu}, X_\mu, P_\mu, h$ are: 
    \begin{align}
        [\Theta_{\mu\nu},\Theta_{\rho\sigma}]&=i\hbar(\eta_{\mu\rho}\Theta_{\nu\sigma}+\eta_{\nu\sigma}\Theta_{\mu\rho}-\eta_{\nu\rho}\Theta_{\mu\sigma}-\eta_{\mu\sigma}\Theta_{\nu\rho})\,,\\
        [P_\mu,P_\nu]&=i\frac{\hbar}{\lambda^2}\Theta_{\mu\nu}\,,\quad\quad [X_\mu,X_\nu]=i\frac{\lambda^2}{\hbar}\Theta_{\mu\nu}\,,\label{definingrelation}\\
        [P_\mu,h]&=-i\frac{\hbar}{\lambda^2}X_\mu\,,\quad\quad [X_\mu,h]=i\frac{\lambda^2}{\hbar}P_\mu\,,\\
        [\Theta_{\mu\nu},P_\rho]&=i\hbar(\eta_{\mu\rho}P_\nu-\eta_{\nu\rho}P_\mu)\,,\quad\quad [\Theta_{\mu\nu},X_\rho]=i\hbar(\eta_{\mu\rho}X_\nu-\eta_{\nu\rho}X_\mu)\,,\\
        [P_\mu,X_\nu]&=i\hbar\eta_{\mu\nu}h\,,\quad\quad [\Theta_{\mu\nu},h]=0\,.
    \end{align}
\item The above algebra, in which we resulted, is identical to the one we used in our previous work \cite{Manolakos:2019fle}. This time it has been derived in a 2-step decomposition procedure, with the results of the first one emphasizing on its origin as an embedding in the 5-dim Minkowski spacetime but also on its form, rendering more apparent that it is indeed a fuzzy version of the 4-dim de Sitter space. After the second step we have expressed the covariant 4-dim fuzzy de Sitter space in $SO(1,3)$ terms and also captured the momentum space as a bonus. Last, the embedding relation written in eq.\eqref{embeddingrelation}, is written in the $SO(1,3)$ language as:
\begin{equation}
    X^m X_m=r^2\,\Rightarrow X^\mu X_\mu+X_4X^4=r^2\Rightarrow h=\pm\sqrt{\frac{1}{\lambda^2}(X_\mu X^\mu-r^2)}\,,
\end{equation}
where as it turns out, $h$ is an operator bearing the information of the radius constraint, namely the embedding equation obtained in \eqref{embeddingrelation}, of the fuzzy space can be interpreted as an expression of $h$ in this $SO(1,3)$ picture.
\end{itemize}
\section{Determining the action and the field equations}
In the context of the noncommutative $SO(2,4)\times U(1)$ gauge theory we wrote down in our previous work \cite{Manolakos:2019fle}, we mean to add some more information about the origins of the considered action and the field equations that are produced. For reasons of self-consistency of the current work, we recall that we started with a gauge theory of the isometry group of our fuzzy space, namely $SO(1,5)$ and extended it to to the $SO(2,4)\times U(1)$ in a fixed representation due to the non-closure property of the anticommutators of the generators. The sixteen generators of the algebra in the  $4\times 4$ representation are the following:\\ 
a) Six Lorentz generators:
$ \mathrm{M}_{ab} =  - \dfrac{i}{4} [\Gamma_{a} , \Gamma_{b} ] = - \dfrac{i}{2} \Gamma_{a} \Gamma_{b}\,,a < b$,\\\\
b) four generators for special conformal transformations: $ \mathrm{K}_{a} = \dfrac{1}{2} \Gamma_{a}$,\\\\
c) four generators for translations: $ \mathrm{P}_{a} = -\dfrac{i}{2} \Gamma_{a} \Gamma_{5}$,\\\\
d) one generator for dilatations: $\mathrm{D} = -\dfrac{1}{2} \Gamma_{5}$ and\\\\
e) one $\mathrm{U(1)}$ generator: $\mathbf{I}_4$,\\
where the well-known $4\times 4$ gamma matrices have been used\footnote{For extended arguments on the fixing of the  representation of the generators see \cite{Manolakos:2019fle}. Also, it is worth-noting that the various generators are written already in an $SO(1,3)$ notation, a form that will be useful regarding the symmetry breaking procedure that will follow.}. These generators satisfy the following commutation and anticommutation relations:\\
\begin{align}
&[ K_{a} , K_{b} ] = i M_{ab}, \ \ \ \ \ [P_{a}, P_{b} ] = i M_{ab} \nonumber \\
&[P_{a}, D ] =i K_{a} , \ \ \ \ \ [K_a,P_b]=i\delta_{ab}D , \ \ \ [K_a,D]=-iP_a \nonumber \\
&[K_{a}, M_{bc} ] = i( \delta_{ac} K_{b} - \delta_{ab} K_{c} ) \nonumber \\
&[P_{a}, M_{bc} ] = i( \delta_{ac} P_{b} - \delta_{ab} P_{c} ) \nonumber \\
&[M_{ab}, M_{cd} ] = i( \delta_{ac} M_{bd} + \delta_{bd} M_{ac} - \delta_{bc}M_{ad} - \delta_{ad}M_{bc} ) \nonumber \\
&[D, M_{ab} ] = 0\label{algebra}
\end{align}
\begin{align}
    \{M_{ab},M_{cd}\}&=\frac{1}{8}\left(\delta_{ac}\delta_{bd}-\delta_{bc}\delta_{ad}\right)\mathbf{I}_4-\frac{\sqrt{2}}{4}\epsilon_{abcd}D\nonumber\\
    \{M_{ab},K_c\}&=\sqrt{2}\epsilon_{abcd}P_d\,,\quad \{M_{ab},P_c\}=-\frac{\sqrt{2}}{4}\epsilon_{abcd}K_d\nonumber\\
    \{K_a,K_b\}&=\frac{1}{2}\delta_{ab}\mathbf{I}_4\,,\quad \{P_a,P_b\}=\frac{1}{8}\delta_{ab}\mathbf{I}_4\,,\quad \{K_a,D\}=\{P_a,D\}=0\nonumber\\
    \{P_a,K_b\}&=\{M_{ab},D\}=-\frac{\sqrt{2}}{2}\epsilon_{abcd}M_{cd}\,.\quad \label{anticomso(4)}
\end{align}
Next, searching for the action of the noncommutative $SO(2,4) \times U(1)$ gauge
theory and keeping in mind the defining relation of the noncommutativity of the coordinates of the fuzzy space \eqref{definingrelation}, let us consider first the following topological action\footnote{Along the lines of \cite{Chatzistavrakidis:2018vfi}, where a topological action was the starting point in the corresponding 3-dim analysis.}:
\begin{equation}
    \mathcal{S}=\mathrm{Tr}\left([X_\mu,X_\nu]-\kappa^2\Theta_{\mu\nu}\right)\left([X_\rho,X_\sigma]-\kappa^2\Theta_{\rho\sigma}\right)\epsilon^{\mu\nu\rho\sigma}\,.\label{topologicalaction}
\end{equation}
Variation of the above action will lead to the corresponding field equations. It is expected that our background space should satisfy the derived equations.
\noindent We consider that, in principle, $X$ and $\Theta$ are independent fields. Thus variation of the above action gives:
\begin{align}
    \delta\mathcal{S}&=2\mathrm{Tr}\left(\delta[X_\mu,X_\nu][X_\rho,X_\sigma]-\kappa^2\delta[X_\mu,X_\nu]\Theta_{\rho\sigma}-\kappa^2\delta\Theta_{\mu\nu}[X_\rho,X_\sigma]+\kappa^4\delta\Theta_{\mu\nu}\Theta_{\rho\sigma}\right)\epsilon^{\mu\nu\rho\sigma}\,.\nonumber
\end{align}
Specifying the variation procedure for each of the independent fields we have:
\begin{align}
    \delta_X\mathcal{S}&=4\mathrm{Tr}\delta X_\mu[X_\nu,[X_\rho,X_\sigma]-\kappa^2\Theta_{\rho\sigma}]\epsilon^{\mu\nu\rho\sigma}=0\,.
    \nonumber
\end{align}
Therefore, the first field equation is:
\begin{equation}
    \epsilon^{\mu\nu\rho\sigma}[X_\nu,[X_\rho,X_\sigma]-\kappa^2\Theta_{\rho\sigma}]=0\,.\label{eom1}
\end{equation}
Also, in turn:
\begin{align}
    \delta_\Theta\mathcal{S}&=2\mathrm{Tr}\delta\Theta_{\mu\nu}\left(-\kappa^2[X_\rho,X_\sigma]+\kappa^4\Theta_{\rho\sigma}\right)\epsilon^{\mu\nu\rho\sigma}=0\,.\nonumber
\end{align}
Therefore, the second field equation is:
\begin{equation}
    \epsilon^{\mu\nu\rho\sigma}([X_\rho,X_\sigma]-\kappa^2\Theta_{\rho\sigma})=0\,.\label{eom2}
\end{equation}
The two field equations, eq. \eqref{eom1}, \eqref{eom2}, are satisfied by the fuzzy space we considered when $\kappa^2=\dfrac{i\lambda^2}{\hbar}$, i.e. it is indeed a solution of the action considered in \eqref{topologicalaction}. It should be noted, that the first field equation, \eqref{eom1}, highlights the fact that the background space is not a dynamical one, while the second one, \eqref{eom2}, manifests the dependence between $\Theta$ and $X$.  

Also, retrospective approach of the background solution of the field equations suggests that, since we already know the noncommutative relation-definition of our space, \eqref{definingrelation}, $\Theta$ and $X$ are related and, therefore, we could use the same action, \eqref{topologicalaction} with the difference that $\Theta=\Theta(X)$, that is to assume that $\Theta$ and $X$ are not independent. Performing the variation in this case, we end up only with the first field equation, \eqref{eom1} (since the \eqref{eom2} gave us the information $\Theta=\Theta(X)$ which now we have already taken for granted), which is again satisfied by the fuzzy space.  

Now, we would like to promote the above action, \eqref{topologicalaction}, to a dynamical one, that is to write it in a form in which the gauge fields of the theory would be contained. There are two ways to achieve that. The first one is the straightforward introduction of the gauge fields as fluctuations of the coordinates in the above field equations, \eqref{eom1}, \eqref{eom2}, that is:
\begin{align}
    &\epsilon^{\mu\nu\rho\sigma}\left([X_\rho+A_\rho,X_\sigma+A_\sigma]-\frac{i\lambda^2}{\hbar}(\Theta_{\rho\sigma}+\mathcal{B}_{\rho\sigma})\right)=0\nonumber\\
    &~~~~~~~~~~~~~~~\epsilon^{\mu\nu\rho\sigma}\left([\mathcal{X}_\rho,\mathcal{X}_\sigma]-\frac{i\lambda^2}{\hbar}\hat{\Theta}_{\rho\sigma}\right)=0\nonumber\\
    &~~~~~~~~~~~~~~~~~~~~~~~~~~\epsilon^{\mu\nu\rho\sigma}\mathcal{R}_{\rho\sigma}=0\,,\label{solution1}
\end{align}
\begin{align}
    &\epsilon^{\mu\nu\rho\sigma}\left[X_\nu+A_\nu,[X_\rho+A_\rho,X_\sigma+A_\sigma]-\frac{i\lambda^2}{\hbar}(\Theta_{\rho\sigma}+\mathcal{B}_{\rho\sigma})\right]=0\nonumber\\
    &~~~~~~~~~~~~~~~\epsilon^{\mu\nu\rho\sigma}\left[\mathcal{X}_\nu,[\mathcal{X}_\rho,\mathcal{X}_\sigma]-\frac{i\lambda^2}{\hbar}\hat{\Theta}_{\rho\sigma}\right]=0\nonumber\\
    &~~~~~~~~~~~~~~~~~~~~~~~~~~\epsilon^{\mu\nu\rho\sigma}[\mathcal{X}_\nu,\mathcal{R}_{\rho\sigma}]=0\,.\label{solution2}
\end{align}
The first field equation is the vanishing of the field strength tensor while the second one can be interpreted as a noncommutative analogue of the second Bianchi identity\footnote{Although they are evident, explicit definitions for the field strength tensor, the covariant coordinate and the covariant noncommutative tensor are given below, after eq.\eqref{lastformaction}.}. 

Nevertheless, for our purposes, in order to obtain the field equations, we need to follow a different path (second way), that is expressing the action in terms of the (curvature) field strength tensor, which will emerge in a rather spontaneous way since we follow a gauge-theoretic approach, but also for two more reasons. The first is that we mean to employ the curvature field strength tensor in order to find resemblances with the commutative case and gain intuition by comparison. The second reason is related to the fact that, shortly, we are performing a spontaneous symmetry breaking in order to reduce the gauge symmetry and this happens in the early stage of the determination of the action. Therefore, the initial action we considered, \eqref{topologicalaction}, can now be written in terms of the field strength tensor, by introducing the fluctuations, i.e. the gauge fields, in it:
\begin{align*}
    \mathcal{S}&=\mathrm{Tr}\mathrm{tr}\epsilon^{\mu\nu\rho\sigma}\left([X_\mu+A_\mu,X_\nu+A_\nu]-\kappa^2(\Theta_{\mu\nu}+\mathcal{B}_{\mu\nu})\right)\left([X_\rho+A_\rho,X_\sigma+A_\sigma]-\kappa^2(\Theta_{\rho\sigma}+\mathcal{B}_{\rho\sigma})\right)\,,
\end{align*}
where a trace over the gauge algebra has been also included. The above action, for the specific $\kappa=i\lambda^2/\hbar$, becomes:
\begin{align}
    \mathcal{S}&=\mathrm{Tr}\mathrm{tr}\left([\mathcal{X}_\mu,\mathcal{X}_\nu]-\frac{i\lambda^2}{\hbar}\hat{\Theta}_{\mu\nu}\right)\left([\mathcal{X}_\rho,\mathcal{X}_\sigma]-\frac{i\lambda^2}{\hbar}\hat{\Theta}_{\rho\sigma}\right)\epsilon^{\mu\nu\rho\sigma}:=\mathrm{Tr}\mathrm{tr}\mathcal{R}_{\mu\nu}\mathcal{R}_{\rho\sigma}\epsilon^{\mu\nu\rho\sigma}\,,\label{lastformaction}
\end{align}
where we have defined:
\begin{itemize}
    \item[-] $\mathcal{X}_\mu=X_\mu+A_\mu$, the covariant coordinate of the noncommutative gauge theory with the introduction of the gauge connection $A_\mu$.
    \item[-] $\hat{\Theta}_{\mu\nu}=\Theta_{\mu\nu}+\mathcal{B}_{\mu\nu}$, the covariant noncommutative tensor, with the inclusion of the 2-form field $\mathcal{B}_{\mu\nu}$.
    \item[-] $ \mathcal{R}_{\mu\nu}=[\mathcal{X}_\mu,\mathcal{X}_\nu]-i\frac{\lambda^2}{\hbar}\hat{\Theta}_{\mu\nu}$, the field strength tensor of the theory\,.
\end{itemize}
The action in its last form \eqref{lastformaction}, is the one we had considered in the first place \cite{Manolakos:2019fle} (the noncommutative analogue of the 4-dim Pontryagin density) and is the one in which we will in turn introduce the scalar field to induce the spontaneous symmetry breaking. Variation of this action with respect to $\mathcal{X}$ and $\mathcal{B}$ leads to the field equations obtained in \eqref{solution1}, \eqref{solution2}:
\begin{equation}
    \epsilon^{\mu\nu\rho\sigma}\mathcal{R}_{\rho\sigma}=0\,,\quad \epsilon^{\mu\nu\rho\sigma}[\mathcal{X}_\nu,\mathcal{R}_{\rho\sigma}]=0\,.\label{bothsolutions}
\end{equation}
Therefore, in this section we extracted two important results:
\begin{itemize}
    \item[-] We defined an action and found that the fuzzy space we formulated is a non-dynamical background solution of the theory. 
    \item[-] Adding fluctuations we obtained the corresponding, dynamical, field equations and an action of Yang-Mills type and extracted the definitions of the covariant coordinate and the field strength tensor of the noncommutative gauge theory in a natural and unforced way.
\end{itemize}
For completeness and later use, we write down the expressions of the component curvature tensors, which are obtained after considering the definition of the field strength tensor, $\mathcal{R}_{\mu\nu}(X)$, and its decomposition on the various generators of the gauge algebra:
\[\mathcal{R}_{\mu\nu}(X)=\tilde{R}_{\mu\nu}^{~~a}\otimes P_a+R_{\mu\nu}^{~~ab}\otimes M_{ab}+R_{\mu\nu}^{~~a}\otimes K_a+ \tilde{R}_{\mu\nu}\otimes D+R_{\mu\nu}\otimes \mathbf{I}_4\,. \]
Therefore, the expressions of the component tensors are:
\begin{align}
R_{\mu\nu}&=[X_\mu,a_\nu]-[X_\nu,a_\mu]+[a_\mu,a_\nu]+\frac{1}{4}[b_\mu^{~a},b_{\nu a}]+\frac{1}{4}[\tilde{a}_\mu,\tilde{a}_\nu]+\frac{1}{8}[\omega_\mu^{~ab},\omega_{\nu ab}]\nonumber\\
&~~~+\frac{1}{16}[e_{\mu a},e_\nu^{~a}]-\frac{i\lambda^2}{\hbar}B_{\mu\nu}\label{R(1)}\\
\tilde{R}_{\mu\nu}&=[X_\mu+a_\mu,\tilde{a}_\nu]-[X_\nu+a_\nu,\tilde{a}_\mu]+\frac{i}{2}\{b_{\mu a},e_\nu^{~a}\}-\frac{i}{2}\{b_{\nu a},e_\mu^{~a}\}\nonumber\\
~~~&-\frac{\sqrt{2}}{8}\epsilon_{abcd}[\omega_\mu^{~ab},\omega_\nu^{~cd}]-\frac{i\lambda^2}{\hbar}\tilde{B}_{\mu\nu}\label{R(D)} \\ 
R_{\mu\nu}^{~~~a}&=[X_\mu+a_\mu,b_\nu^{~a}]-[X_\nu+a_\nu,b_\mu^{~a}]+i\{b_{\mu b},\omega_\mu^{~ab}\}-i\{b_{\nu b},\omega_\mu^{~ab}\}\nonumber \\
&\quad -\frac{i}{2}\{\tilde{a}_\mu,e_\nu^{~a}\}+\frac{i}{2}\{\tilde{a}_\nu,e_\mu^{~a}\}+\frac{\sqrt{2}}{8}\epsilon_{abcd}([e_\mu^{~b},\omega_\nu^{~cd}]-[e_\nu^{~b},\omega_\mu^{~cd}])-\frac{i\lambda^2}{\hbar}B_{\mu\nu}^{~~~a}\label{R(K)}
\end{align}
\begin{align}
\tilde{R}_{\mu\nu}^{~~~a}&=[X_\mu+a_\mu,e_\nu^{~a}]-[X_\nu+a_\nu,e_\mu^{~a}]-\frac{i}{2}\{b_\mu^{~a},\tilde{a}_\nu\}+\frac{i}{2}\{b_\nu^{~a},\tilde{a}_\mu\}\nonumber\\
&\quad -\frac{\sqrt{2}}{2}([b_\mu^{~b},\omega_\nu^{~cd}]-[b_\nu^{~b},\omega_\mu^{~cd}])\epsilon_{abcd}-i\{\omega_\mu^{~ab},e_{\nu b}\}+i\{\omega_\nu^{~ab},e_{\mu b}\}-\frac{i\lambda^2}{\hbar}\tilde{B}_{\mu\nu}^{~~~a}\label{R(P)}\\
R_{\mu\nu}^{~~~ab}&=[X_\mu+a_\mu,\omega_\nu^{~ab}]-[X_\nu+a_\nu,\omega_\mu^{~ab}]+\frac{i}{2}\{b_\mu^{~a},b_\nu^{~b}\}+\frac{\sqrt{2}}{4}([b_\mu^{~c},e_\nu^{~d}]-[b_\nu^{~c},e_\mu^{~d}])\epsilon_{abcd}\nonumber \\
&\quad -\frac{\sqrt{2}}{4}([\tilde{a}_\mu,\omega_\nu^{~cd}]-[\tilde{a}_\nu,\omega_\mu^{~cd}])\epsilon_{abcd}+2i\{\omega_\mu^{~ac},\omega_{\nu~c}^{~b}\}+\frac{i}{2}\{e_\mu^{~a},e_\nu^{~b}\}-\frac{i\lambda^2}{\hbar}B_{\mu\nu}^{~~~ab}\label{R(M)}\,.
\end{align}
\section{Spontaneous symmetry breaking of the noncommutative action}\label{noncommutativeconformalbreaking}
According to our previous work \cite{Manolakos:2019fle}, in which we constructed a 4-dim noncommutative gravity model as a gauge theory of the $G=SO(2,4)\times U(1)$, we considered the following action\footnote{To be precise, in our previous work \cite{Manolakos:2019fle}, the action we considered was an extension of this one, namely:
\[\mathcal{S}=\text{Tr}\text{tr}_G\,\Gamma_5(\mathcal{R}_{\mu\nu}\mathcal{R}_{\rho\sigma}\varepsilon^{\mu\nu\rho\sigma}+\hat{H}_{\mu\nu\rho}\hat{H}^{\mu\nu\rho})\,,\] where the second term is the kinetic term of the $\mathcal{B}$ field that had been introduced. In the present study, we omit this term as we regard the $\mathcal{B}$ field to be non-propagating.}:
\begin{equation}
    \mathcal{S}=\text{Tr}\text{tr}_G\,\Gamma_5\mathcal{R}_{\mu\nu}\mathcal{R}_{\rho\sigma}\varepsilon^{\mu\nu\rho\sigma}\,,\label{actionprevious}
\end{equation}
which is actually the action we formed in \eqref{lastformaction}, after the introduction of fluctuations in the non-dynamical starting action \eqref{topologicalaction}, accompanied by the matrix $\Gamma_5$. The latter was used in order that we could deliver the symmetry breaking, which was realized by the imposition of certain constraints, to the action and result with an action that would include an $R(M)^2$ term, where $R(M)$ is the component curvature 2-form of the total field strength $\mathcal{R}$ and is associated to the Lorentz generators, see eq.\eqref{R(M)}. 

This time we intend to perform a spontaneous symmetry breaking with the inclusion of a scalar field, $\Phi$, which belongs to the adjoint representation of $SO(2,4)\times U(1)$. Therefore, our starting point is exactly the action we obtained in the previous section, namely  eq. \eqref{lastformaction}, that is the above one dropping the $\Gamma_5$ matrix: 
\begin{equation}
    \mathcal{S}_0=\text{Tr}\text{tr}_G\,\mathcal{R}_{\mu\nu}\mathcal{R}_{\rho\sigma}\varepsilon^{\mu\nu\rho\sigma}\,.\label{neataction}
\end{equation}
The above action can be written in the following equivalent form, that is a modification induced by the introduction of the scalar field along with the dimensionful parameter, $\lambda$, which is the length scale of our theory\footnote{The parameter $\lambda$ is the length dimension ($[\lambda]=L^1=M^{-1}$) we introduced during the procedure of the construction of the covariant fuzzy space.}:
\begin{equation}
    \mathcal{S}=\text{Tr}\text{tr}_G\, \lambda\Phi(X)\mathcal{R}_{\mu\nu}\mathcal{R}_{\rho\sigma}\varepsilon^{\mu\nu\rho\sigma}+\eta(\Phi(X)^2-\lambda^{-2}\mathbf{I}_N\otimes\mathbf{I}_4)\,,\label{actionwithscalars}
\end{equation}
where $\eta$ is a Lagrange multiplier. The initial form of the action, \eqref{neataction}, is recovered if the following constraint equation of $\eta$ is taken into account (on-shell):
\begin{equation}
   \Phi^2(X)=\lambda^{-2}\mathbf{I}_N\otimes\mathbf{I}_4\,.
\end{equation}
The Lagrange multiplier, $\eta$ is of dimension $[M]^2$ and variation of the above action with respect to it, produces the above constraint of the auxiliary field as equation of motion. Also, since the field $\Phi$ belongs to the gauge algebra, in principle, it can be written as a decomposition on the sixteen generators:
\begin{align}
    \Phi(X)&=\tilde{\phi}^a(X)\otimes P_a+\phi^{ab}(X)\otimes M_{ab}+\phi^a(X)\otimes K_a+\phi(X)\otimes\mathbf{I}_4+\tilde{\phi}(X)\otimes D\,. \label{phi}
\end{align}
We return to the action of eq.\eqref{actionwithscalars} in which we have to calculate the product of the first term. Due to the insight given to us by our previous work and since we desire to result with an $R(M)^2$ term, we begin with the $\Phi$ field and gauge fix it in the direction of the generator $D$, which will play the role of the $\Gamma_5$ of the action of eq.\eqref{actionprevious} introduced in a more natural way this time (reminding that $D=-1/2\Gamma_5$), at the specific value of $\tilde{\phi}=-2\lambda^{-1}$. Therefore, making use of the decomposition of $\Phi$, \eqref{phi}, we write down the gauge fixed expression\footnote{We are motivated to concentrate on gauge fixing in the $D$ direction of the field since the breaking the scale invariance suffices to break the special conformal symmetry as well.}:
\begin{align}
    \Phi(X)=\tilde{\phi}(X)\otimes D|_{\tilde \phi=-2\lambda^{-1}}=-2\lambda^{-1}\mathbf{I}_N\otimes D\,.
\end{align}
Taking into consideration the anticommutation relations of the various generators of the group, \eqref{anticomso(4)}, the various traces over the algebra are calculated. The surviving contributions of the action in \eqref{actionwithscalars}, after the gauge fixing, constitute the following form of the spontaneously broken action:
\begin{equation}
    \mathcal{S}_{br}=\text{Tr}\left(\frac{\sqrt{2}}{4}\varepsilon_{abcd}R_{mn}^{~~ab}R_{rs}^{~~cd}-4R_{mn}\tilde{R}_{rs}\right)\varepsilon^{mnrs}\,,\label{actionafterbreaking}
\end{equation}
where it is clear that the Lagrange multiplier term of \eqref{actionwithscalars} totally vanishes in this gauge. It should be noted that the action we ended up coincides with the one we presented in our previous work obtained through different arguments \cite{Manolakos:2019fle}.



Now, motivated by A. Chamseddine's work \cite{Chamseddine:2002fd}, we set $\tilde a_m=0$ and $b_m^{~a}=\alpha e_m^{~a}$ (along with $B_{mn}^{~~a}=\alpha \tilde B_{mn}^{~~a}$), where $\alpha$ is a proportionality constant. We also assume the torsionless condition, since the corresponding (translational) part of the symmetry has been broken, as the corresponding tensor has been dropped out of the action. The latter gives a relation of $\omega$ with respect to the independent fields. Letting $\tilde a_m=0$ and $b, e$ to be proportional to each other, it is straightforward to observe, through their corresponding expressions \eqref{R(K)}, \eqref{R(P)}, that if the constant of proportionality is $\alpha=\frac{i}{2}$, then the $K$-tensor, $R_{mn}^{~~a}$, is equal to the torsion tensor, $\tilde R_{mn}^{~~a}$, up to the same constant of proportionality:
\begin{align}
    R_{mn}^{~~a}&=[X_m+a_m,b_m^{~a}]-[X_n+a_n,b_m^{~a}]+i\{b_{mb},\omega_n^{~ab}\}-i\{b_{nb},\omega_m^{~ab}\}\nonumber\\
    &~~+\frac{\sqrt{2}}{8}\epsilon_{abcd}([e_m^{~b},\omega_n^{~cd}]-[e_n^{~b},\omega_m^{~cd}])-i\frac{\lambda^2}{\hbar} B_{mn}^{~~a}\nonumber \\
    &=\frac{i}{2}\left([X_m+a_m,e_n^{~a}]-[X_n+a_n,e_m^{~a}]+i\{e_{mb},\omega_n^{~ab}\}-i\{e_{nb},\omega_m^{~ab}\}\right.\nonumber\\
    &~~\left.-\frac{\sqrt{2}}{2}\epsilon_{abcd}([b_m^{~b},\omega_n^{~cd}]-[b_n^{~b},\omega_m^{~cd}])-i\frac{\lambda^2}{\hbar}\tilde B_{mn}^{~~a}\right)=\frac{i}{2}\tilde R_{mn}^{~~a}\,.\nonumber
\end{align}
Therefore, we conclude that the $K-$related field strength tensor is also vanishing, namely $R_{mn}^{~~a}=0$, which means that the corresponding generators have been broken, too (for the specific gauge $\tilde a_m=0$, $b_m^{~a}=\frac{i}{2}e_m^{~a}$ and $B_{mn}^{~~a}=\frac{i}{2}\tilde{B}_{mn}^{~~a}$). 

Before we move on, this is a nice point to examine the torsionless condition  which leads to the relation between the $\omega$ field with respect to the independent field $e$ (since $\tilde{a}$ has been set equal to zero):
\begin{align}
    \tilde R_{mn}^{~~a}&=[X_m+a_m,e_n^{~a}]-[X_n+a_n,e_m^{~a}]-\frac{\sqrt{2}i}{4}([e_m^{~b},\omega_n^{~cd}]-[e_n^{~b},\omega_m^{~cd}])\epsilon_{abcd}\nonumber \\
    &~~-i\{\omega_m^{~ab},e_{nb}\}+i\{\omega_n^{~ab},e_{mb}\}-i\frac{\lambda^2}{\hbar}\tilde B_{mn}^{~~a}=0\,.\label{torsionless}
\end{align}
We will see that the above relation (torsionless condition) will reduce to that of the Einstein-Hilbert case when the commutative limit will be applied. 

Moreover, we find the explicit expression of the $R_{mn}^{~~ab}$ component of the field strength tensor which is present in the above action, \eqref{actionafterbreaking}, and will do survive in the commutative regime. Applying the gauge fixing conditions on the fields, $\tilde a_m=0, b_m^{~a}=\frac{i}{2}e_m^{~a}$, we get:
\begin{align}
    R_{mn}^{~~ab}&=[X_m+a_m,\omega_n^{~ab}]-[X_n+a_n,\omega_m^{~ab}]+i\{\omega_m^{~ac},\omega_{n~c}^{~b}\}-i\{\omega_m^{~bc},\omega_{n~c}^{~a}\}\nonumber\\
    &~~+\frac{3i}{8}\{e_m^{~a},e_n^{~b}\}-\frac{i\lambda^2}{\hbar}B_{mn}^{~~ab}\,.\label{curvature}
\end{align}
In the consideration of the commutative limit, it will be clear that the first line of the above equation will be identical to the $R_{mn}^{(0)ab}$, that is the curvature 2-form in the Palatini formulation of General Relativity. 

\section{The commutative limit after the symmetry breaking}

In order to examine the implications of our noncommutative gravity model in the energy regime below the Planck scale, we need to consider the commutative limit. In this limit, the fuzzy space reduces to the ordinary 4-dim de Sitter space and, in order to obtain the corresponding gauge theory, we make the following considerations:
\begin{itemize}
    \item The 2-form field, $\mathcal{B}_{\mu\nu}$, that is related to the preservation of the covariance of the field strength tensor in the noncommutative regime, decouples as the noncommutativity of the space ceases to exist. The same applies for the $a_\mu$ field\footnote{A possible mechanism for these two fields, which are strongly related to the noncommutative nature of the spacetime in the high-energy regime, to decouple may be related to a mass-gaining mechanism which is yet to be examined.}.
    \item The commutators of functions vanish: $[f(x),g(x)]\,\rightarrow\,0$
    \item The anticommutators of functions reduce to product: $\{f(x),g(x)\}\,\rightarrow\,2f(x)g(x)$
     \item The inner derivation becomes: $[X_\mu,f]\,\rightarrow\,\partial_\mu f$
    \item Trace reduces to integration: $\dfrac{\sqrt{2}}{4}\mathrm{Tr}\,\rightarrow \int d^4x$
    \item From \eqref{R(D)}, it is easy to see that, in the specific gauge, in which the symmetry breaking occurs, the $\tilde{R}_{mn}$ tensor takes the following form:
    \begin{equation}
        \tilde{R}_{\mu\nu}=-\frac{\sqrt{2}}{8}\epsilon_{abcd}[\omega_\mu^{~ab},\omega_\nu^{~cd}]-\frac{i\lambda^2}{\hbar}\tilde{B}_{\mu\nu}\,.
    \end{equation}
    Therefore, consideration of the commutative limit will lead to the vanishing of the second term of the corresponding action, \eqref{actionafterbreaking}, which involves the above $\tilde{R}_{\mu\nu}$ tensor, as $\tilde{B}_{\mu\nu}$ decouples and the commutator of the spin connection functions is zero. Also, in this limit, the first term of the action \eqref{actionafterbreaking}, which contains the tensor $R_{\mu\nu}^{~~ab}$, ceases to include the $a_\mu$ field, since it also decouples\footnote{It is also easy to see that it is involved only in commutators with other fields which eventually all vanish.}.
    \item We also regard the following reparametrizations:
    \item[-] $e_\mu^{~a}\,\rightarrow\, ime_\mu^{~a}\,,\quad P_a\,\rightarrow\,-\dfrac{i}{m}P_a\,,\quad \tilde{R}_{\mu\nu}^{~~a}\,\rightarrow\, imT_{\mu\nu}^{~~a}$
    \item[-] $\omega_\mu^{~ab}\,\rightarrow\, -\dfrac{i}{2}\omega_\mu^{~ab}\,,\quad M_{ab}\,\rightarrow\,2iM_{ab}\,,\quad {R}_{\mu\nu}^{~~ab}\,\rightarrow\, -\dfrac{i}{2}R_{\mu\nu}^{~~ab}$,\\
in order to exactly match the results of the commutative case. The $m$ is an arbitrary (complex) constant of dimension $[L]^{-1}$ which is imported in order that the $e_\mu^{~a}$ in the commutative limit to be dimensionless, as it must, in order to admit the interpretation of the actual vielbein.
\end{itemize}

First of all, let us deal with the torsion given in \eqref{torsionless}. Taking the above limits and reparametrizations into consideration we get:
\begin{align}
    imT_{\mu\nu}^{~~a}&=im\partial_\mu e_\nu^{~a}-im\partial_\nu e_\mu^{~a}-2i\left(-\frac{i}{2}\omega_\mu^{~ab}\right)ime_{\nu b}+2i\left(-\frac{i}{2}\omega_\nu^{~ab}\right)ime_{\mu b}\,\Rightarrow \nonumber\\
    T_{\mu\nu}^{~~a}&=\partial_\mu e_\nu^{~a}-\partial_\nu e_\mu^{~a}-\omega_\mu^{~ab}e_{\nu b}+\omega_\nu^{~ab}e_{\mu b}=0\,.
\end{align}
Therefore, in the commutative limit, the torsionless condition is the same as that of the first-order formulation of General Relativity. Thus, we understand that the expression which relates the $\omega$ to the $e$ is exactly the same. 

Let us now examine the expression of the curvature 2-form. Starting with \eqref{curvature} and taking the limits we have:
\begin{align}
    -\frac{i}{2}R_{\mu\nu}^{~~ab}&=-\frac{i}{2}\partial_\mu\omega_\nu^{~ab}+\frac{i}{2}\partial_\nu\omega_\mu^{~ab}+2i\left(-\frac{i}{2}\omega_\mu^{~ac}\right)\left(-\frac{i}{2}\omega_{\nu~ c}^{~b}\right)\nonumber\\
    &~~-2i\left(-\frac{i}{2}\omega_\mu^{~bc}\right)\left(-\frac{i}{2}\omega_{\nu ~c}^{~a}\right)+\frac{3i}{4}(ime_\mu^{~a}ime_\nu^{~b})\Rightarrow\nonumber\\
    R_{\mu\nu}^{~~ab}&=\partial_\mu\omega_\nu^{~ab}-\partial_\nu\omega_\mu^{~ab}+\omega_\mu^{~ac}\omega_{\nu~c}^{~b}-\omega_\mu^{~bc}\omega_{\nu~c}^{~a}+\frac{3}{2}m^2e_\mu^{~a}e_\nu^{~b}\Rightarrow \nonumber\\
    R_{\mu\nu}^{~~ab}&=R_{\mu\nu}^{(0)ab}+\frac{3}{2}m^2e_\mu^{~a}e_\nu^{~b}\,.
\end{align}
In the last relation it is emphasized that the curvature 2-form in our case is the one of the General Relativity case in its first order formulation, along with an extra term which will contribute in getting the Einstein-Hilbert action besides the Gauss-Bonnet topological term. 

Last, as we remarked above, applying the limits leads to the vanishing of the second term of the action \eqref{actionafterbreaking}. Therefore, the action consists only of the first $R(M)^2$ term and, given that the scale invariance has been broken spontaneously, the introduction of a dimensionful parameter, $m$, into the theory causes no problems. Therefore, the only invariance the action enjoys in the commutative limit is the Lorentz. Let us move on with the calculations and examine what is produced:
\begin{align}
    \mathcal{S}_{br}^{comm}&=\int 
    \epsilon_{abcd}R_{\mu\nu}^{~~ab}R_{\rho\sigma}^{~~cd}\epsilon^{\mu\nu\rho\sigma}d^4x\nonumber\\
    &=\int \epsilon_{abcd}\left(R_{\mu\nu}^{(0)ab}+\frac{3}{2}m^2e_\mu^{~a}e_\nu^{~b}\right)\left(R_{\rho\sigma}^{(0)cd}+\frac{3}{2}m^2e_\rho^{~c}e_\sigma^{~d}\right)\epsilon^{\mu\nu\rho\sigma}d^4x\nonumber\\
    &=\int \epsilon_{abcd}R_{\mu\nu}^{(0)ab}R_{\rho\sigma}^{(0)cd}\epsilon^{\mu\nu\rho\sigma}d^4x+3m^2\int \epsilon_{abcd}e_\mu^{~a}e_\nu^{~b}R_{\rho\sigma}^{(0)cd}\epsilon^{\mu\nu\rho\sigma}d^4x\nonumber\\
    &~~~+\frac{9}{4}m^4\int\epsilon_{abcd}e_\mu^{~a}e_\nu^{~b}e_\rho^{~c}e_\sigma^{~d}\epsilon^{\mu\nu\rho\sigma}d^4x\,.\nonumber
\end{align}
The first term is exactly the topological Gauss-Bonnet term, therefore we drop it since it does not contribute to the equations of motion. The second term is the Palatini action, which is an alternative to the Einstein-Hilbert action, and the last one is a cosmological constant term. After some calculations we have:
\begin{align}
    \mathcal{S}_{br}^{comm}&=12m^2\left(\int\sqrt{\mathrm{det}g}R\,d^4x+\frac{9m^2}{2}\int \sqrt{\mathrm{det}g}\,d^4x\right)\,.
\end{align}
Redefining: $\Lambda=-\frac{9}{4}m^2$ the above action becomes:
\[\mathcal{S}_{br}^{comm}=12m^2\mathcal{S}_{EH}^{(\Lambda)}\,,\] where $\mathcal{S}_{EH}^{(\Lambda)}$ is the 4-dim Einstein-Hilbert action with cosmological constant and without matter. Variation of the above action, leads to the expected Einstein's field equations:
\[ R_{\mu\nu}-\frac{1}{2}R g_{\mu\nu}+\Lambda g_{\mu\nu}=0\,.\]
Assuming that $\Lambda$ is positive, the solution of the above equation is the maximally symmetric space with positive curvature, that is the 4-dim de Sitter space we had already considered as the background space, confirming the validity of the consideration. This does not come as a surprise, since in the gauge-theoretic approaches of gravitational theories, the background space is already determined from the start in order to accommodate the gauge theory. The space is a solution of the Einstein equation.

\section{Outlook and Conclusions}

In this work we presented an extension of our previous work \cite{Manolakos:2019fle}, in which we started a programme of the description of a gravitational theory as a gauge theory on a covariant noncommutative space. In particular, we formulated a fuzzy version of the 4-dim de Sitter space and then we constructed an $SO(2,4)\times U(1)$ gauge theory on it, finding the transformations of the various fields introduced and the expressions of their corresponding component field strength tensors. Eventually, we proposed an action and went on with a symmetry breaking with the imposition of certain constraints. 

In the present work, we have elaborated the construction of our fuzzy space in more detail, formulating it in a 2-step procedure which gives rise to more properties, but also its correlation with its non-fuzzy analogue gets highlighted. In turn, we started with an action confirming that the background space we defined is actually a vacuum solution and then we produced its dynamical form by introducing fluctuations on the coordinates. The produced action is the one we had considered in our previous work, namely $SO(2,4)\times U(1)$ invariant. Also, the form of the field strength tensor emerged in a natural way. Finally, we introduced an auxiliary scalar field in the action and proceeded with a spontaneous symmetry breaking due to the scalar field acquiring a vacuum expectation value. The resulting action is invariant under the subgroup $SO(1,3)\times U(1)$ symmetry. With respect to the commutative limit, we ended up with the Einstein's field equations in the case that there is no matter and with the presence of the cosmological constant.

It is rather interesting that we started with a gravitational model in a regime where the noncommutative geometry is the appropriate framework to describe the background spacetime (Planck scale) and resulted with the well-defined Einstein's gravitational theory in the commutative limit. It should be remarked that although the cosmological constant had not been considered in the initial theory, it emerged at low energies (commutative limit). Also, it should be noted that the spontaneous symmetry breaking in the noncommutative theory broke a large part of the initial symmetry, leading to an $SO(1,3)\times U(1)$ gauge symmetry. In the commutative limit the theory is only Lorentz invariant, as expected.  

From now on, having established a well-described gravitational model in the noncommutative framework and having connected it to a solid ordinary gravitational theory, we may proceed with studying its cosmological consequences.

\noindent \\
\textbf{Acknowledgements}:\\
We would like to thank Maja Buri\'c, Ali Chamseddine, Dumitru Ghilencea, Ichiro Oda, Emmanuel Saridakis, Athanasios Chatzistavrakidis and Harold Steinacker for useful discussions. All authors have been supported by the Basic Research Programme, PEVE2020 of National Technical University of Athens, Greece. 
One of us (GZ) would like to thank the DFG Exzellenzcluster 2181:STRUCTURES
of Heidelberg University, MPI and CERN-TH for support. The work of GM is supported by the Croatian Science Foundation project IP-2019-04-4168.

\end{document}